\documentclass[aps,prl,amsmath,amssymb,onecolumn,nofootinbib,showpacs,tightenlines,12pt]{revtex4}

\usepackage{epsfig,graphicx}
\usepackage{bm}
\newcommand{\be}{\begin{equation}}
\newcommand{\ee}{\end{equation}}
\newcommand{\bea}{\begin{eqnarray}}
\newcommand{\eea}{\end{eqnarray}}

\begin{document}
\title{Exact solution for the massless Cylindrically Symmetric Scalar Field in General Relativity, with   Cosmological Constant }
\author{D.Momeni\footnote{Corresponding author}}
\email{davood.momeni@kiau.ac.ir}
 \affiliation{Department of Physics,Azad University, P.O. Box  31485-313, Karaj, Iran}
\author{H. Miraghaei }
\email{ miraghaee @ physics.sharif.ir }
 \affiliation{Department of Physics, Sharif University of Technology, P.O. Box 11365–9161, Tehran, Iran }
\begin{abstract}
In this article we present a new exact solution for scalar field
with cosmological constant in  cylindrical symmetry. Associated
cosmological models  is discussed. One of them describes a Cyclic
universe.
\end{abstract} \maketitle
\section{\label{sec:level1}Introduction}
 In 1959 Buchdahl introduced a new generating method based
 on an old fascinated method which arisen from potential theory
 (reciprocal transformation) for generating exact solution from a
 known perfect fluid solution to another one  \cite{1}. It's family divided
 into two subclass which in one, scalar field depends only
 on  radial coordinates (Cylindrical or Spherical ) and in the other
 -as Wyman mentioned- scalar fields depend only on time \cite{2}.
 For a gravity which is coupled to a neutral, massless scalar field,
  Wyman suggested a method of solution in power series valid for
  a scalar field depends only on time.
   Wyman's static spacetime with a non static
    scalar field could not integrated explicitly  as we noted
     it in \cite{3}. Another generalizations of
 this family to higher dimensional cases could be found in \cite{4}. Recently Vuille found a new exact
 solution for a massless scalar field with cosmological constant in
 plane symmetry \cite{6}. Late  Anzhong Wang studied cylindrically symmetric spacetimes with
  homothetic self-similarity  in the context of Einstein's Theory of Gravity  \cite{9}. Also  Sarioglu  and Tekin
   generalize the four dimensional cylindrically symmetric static
 spacetimes with a negative cosmological constant to higher
dimensions\cite{17}. There is no general solution for Buchdahl
massless scalar field which is minimally coupled
 to a massive one or in the presence of cosmological constant.
 In this letter ,First we present a new cylindrically symmetric solution for
 a massless scalar field coupled minimally to gravity but contains
 cosmological constant. This exact solution is regular for all values of scalar field parameter $\xi^2>0$
 . The purely complex value of $\xi$ contradicts with our assumption
 that the scalar field function must be real.But this complex value
 for $\xi$ imposes a  naked singularity .
 Then we show this solution represents Buchdahl solution in the absence of
 the cosmological constant and reduces to Levi-Civita-$\Lambda$  solution ($LC\Lambda$)  when the
 scalar field set to be zero.Thus we generalized both Buchdahl and
 $LC\Lambda$ solutions and  we solved an old  obscure problem since
 1959.Also we describe their cosmological metrics and show that
 there are two different family of models :one which is simpler
 corresponds to an inflationary model that the scalar field
 has the role of inflaton. The Hubble parameter of this model
 remains constant. Hence this model has no time evolution. This
 solution matches a negative cosmological constant which is
 harmful for accelerating expansion of universe in now era.Another
 solution describes a Friedmann-Robertson-Walker universe with a period of about cosmic
 time scale.This solution related to a Cyclic model
 which has a significant braneworld scenario of the universe.
\section{Field equations}
 We begin with a general cylindrically symmetric metric
in Weyl coordinates   $(t,r,\varphi,z)$, \footnote{I will mostly use
natural units $\hbar = c = 1$  and $8\pi G = 1$.}
\begin{eqnarray}
ds^{2}=-e^{u(r)}dt^2+dr^2+e^{v(r)}d\varphi^2+e^{w(r)}dz^2
\end{eqnarray}
Field equation for a massless minimally coupled scalar field in the
presence of a cosmological constant term $\Lambda$ is reading as:
\begin{eqnarray}
R_{\mu\nu}-\Lambda g_{\mu\nu}=\phi_{;\mu}\phi_{;\nu}
\end{eqnarray}
 We labled metric functions as $u_{i}=\{u(r),v(r),w(r)\},\phi\equiv \phi(r)$ , $\acute{f}=\frac{d f}{d
 r}$.
 In terms of these functions we can rewrite field equation (2) in the
 following succinct forms:
\begin{eqnarray}
2u_{i}''+u_{i}'\sum^{3}_{j=1}u_{j}'-4 \Lambda=0, i=\{1,2,3\}\\
2\sum^{3}_{j=1}u_{j}''+\sum^{3}_{j=1}u_{j}'^{2}-4 \Lambda=4\phi'^{2}
\end{eqnarray}
This system of differential equations  has no simple solution and
only could be reduced to a set of  non linear Sturm-Liouvile
equations.
\section{Exact solutions}
In this section we present some different possible solutions of
(3,4).
\subsection*{Case (1) : solution with $u(r)=constant,w(r)=constant$,$\phi'=0$}
 We prove that if we have $u(r),w(r)=constant$ then we deduce flat
space. Starting from the equation  (3) for $i=1,3$ we get
$\Lambda=0$ and by comparing it with (4) we observe that $\phi=c$ (a
constant) which we can take it zero.We have:
\begin{eqnarray}\nonumber
2v''+v'^{2}=0
\end{eqnarray}
The exact solution for this odinary differential equation
is\footnote{$\log(x)=\int^{x}_{1}\frac{d \zeta}{\zeta}$}:
\begin{eqnarray}\nonumber
v(r)=2\log(a r+b)
\end{eqnarray}
 The constant  $b$ may be turned to zero by changing scales
along the $t$ and $z$ axes and choosing the zero point of the $r$
coordinate. Then the line element (1) transformed to;
\begin{eqnarray}\nonumber
ds^{2}=-dt^2+dr^2+(a r)^2d\varphi^2+dz^2
\end{eqnarray}
 By applying change of coordinates:
\begin{eqnarray}\nonumber
\tilde{\varphi}=a\varphi
\end{eqnarray}
Finally we obtain:
\begin{eqnarray}\nonumber
ds^{2}=-dt^2+dr^2+r^2d\tilde{\varphi}^2+dz^2
\end{eqnarray}
Which obviously is flat space (locally)in cylindrical Weyl
coordinates. We mention here that the flat space obtained here has a
conical nature. As we noted in \cite{14} the conical parameter $a$
related to the gravitational mass per unit length of the spacetime,
$\eta$ , as
\begin{eqnarray}\nonumber
a=1-4\eta
\end{eqnarray}
Such that  $0 < a < 1$. This metric, exposing the geometry around a
straight cosmic string, which is locally identical flat spacetime
however it is not globally Euclidian, since the angle
$\tilde{\varphi}=a\varphi$ varies in the range $0<\tilde{\varphi}<B$
where $b=2\pi a=2\pi(1-4\eta)$.

\subsection*{Case (2) : solution with $u(r)=v(r)=w(r)\neq constant$,$\phi'\neq0$}
 We must mention that the system of differential equations  (3,4)
posses a special solution  $\{u_{i}(r)\}=u(r)$ for the following
equations:
\begin{eqnarray}
2u''+3u'^{2}-4 \Lambda=0\\
6u''+3u'^{2}-4 \Lambda=4\phi'^{2}
\end{eqnarray}
Immediately we can obtain an integral for the field:
\begin{eqnarray}
\phi=\pm\int\sqrt{u''}  dr
\end{eqnarray}
The differential equation (5) is a non linear  2'nd order,and posses
the general solution   for metric function $u(r)$ :
\begin{eqnarray}
u(r)=\pm2\sqrt{\frac{\Lambda}{3}}r+\frac{1}{3}\log[\frac{3}{16\Lambda}(c_{1}e^{2\sqrt{3\Lambda}r}-c_{2})^2]
\end{eqnarray}
The differential equation (5) is solved by taking $ u(r)=a r+y(r)$
,matching constant value  $a$  and  solving differential equation
  $2y''+y'^{2}-4 \Lambda y^2=0 $  or by a simple change of variable from
$u(r)$ to $p=u'$ and consequently rewriting (5) to a first order differential equation for $p(u)$.\\
 Substituting (8) in (7) we obtain:
\begin{eqnarray}
\phi(r)=\mp2 i
\frac{\sqrt{6}}{3}\tanh^{-1}(\sqrt{\frac{c_{1}}{c_{2}}}e^{\sqrt{3\Lambda}r})
\end{eqnarray}
Substituting (8) in (1) and calculating Ricci scalar we obtain:
\begin{eqnarray}
R=4\Lambda[\frac{c_{2}^{2}+c_{1}^{2}e^{4\sqrt{3\Lambda}r}-4c_{1}c_{2}e^{2\sqrt{3\Lambda}r}}{(c_{1}e^{2\sqrt{3\Lambda}r}-c_{2})^2}]
\end{eqnarray}
Contraction the field equation (2) and substituting metric function
$u$ ,with the field solution  gives the same result. Scalar field
must be a real function, then it is proper to define a new parameter
as:
\begin{eqnarray}\nonumber
\sqrt{\frac{c_{1}}{c_{2}}}=i\xi
\end{eqnarray}
 Which in term of $\xi$ the metric and  field functions (8,9) transform to
 functions:
\begin{eqnarray}
u(r)=\pm2\sqrt{\frac{\Lambda}{3}}r+\frac{1}{3}\log[(\xi^2e^{2\sqrt{3\Lambda}r}+1)^2]
\\
\phi=\pm2 \frac{\sqrt{6}}{3}\tan^{-1}(\xi e^{\sqrt{3\Lambda}r})
\end{eqnarray}
We attend that only the minus sign in metric function is consistent
with scalar field equation of motion  $\Box\phi=0$ and the positive
signs makes it with wrong conclusion \footnote{If one write the
 equation of motion for a massless minimally coupled scalar field in a metric like (14)
 $ds^{2}=dr^2+f(r)(-dt^2+d\varphi^2+dz^2)$ we observe that $f^{3/2}(r)\phi'=c.t.e$. By
 writing $f(r)=e^{\pm2\sqrt{\frac{\Lambda}{3}}r}(\xi^2e^{2\sqrt{3\Lambda}r}+1)^{2/3}$ and
  $\phi=\pm2 \frac{\sqrt{6}}{3}\tan^{-1}(\xi e^{\sqrt{3\Lambda}r})$
 obviously this statement is proved. }. Remind that we take
$c_{2}=4\sqrt{\frac{\Lambda}{3}}$ to eliminate the constant term.
Also Ricci scalar (10) can be written as:
\begin{eqnarray}
R=4\Lambda[\frac{1+\xi^{4}e^{4\sqrt{3\Lambda}r}+4\xi^2e^{2\sqrt{3\Lambda}r}}{(\xi^2e^{2\sqrt{3\Lambda}r}+1)^2}]
\end{eqnarray}
And the final form of metric (1) using equation (11) is:
\begin{eqnarray}
ds^{2}=dr^2+e^{-2\sqrt{\frac{\Lambda}{3}}r}(\xi^2e^{2\sqrt{3\Lambda}r}+1)^{2/3}(-dt^2+d\varphi^2+dz^2)
\end{eqnarray}
This is the main result of our article. We obtained an exact
solution which contains two parameter: Cosmological constant term
$\Lambda$ and a parameter $\xi$ associated directly to the scalar
field. Indeed  the constant $\xi$ determines how strongly the scalar
field creates curvature.Clearly the above metric permits $1$
time-like Killing fields $ \frac{\partial}{\partial t}$, which
denotes that the present solution is static, and $3$ space-like
Killing fields (including $2$ translational fields and $1$
rotational ones), which span a Euclidean group  $G(3)$ .The
cosmological horizon(s) for cylindricall symmetric spacetimes is
discussed completely by Wang \cite{18}. The no singularity structure
of the solution (14) is apparent from its Kretschmann scalar.It
reads
\begin{eqnarray}\nonumber
R_{\mu\nu\alpha\beta}R^{\mu\nu\alpha\beta}=\frac{3}{8}[(\frac{f'}{f})^4+2(\frac{f^{''}}{f})^2-2(\frac{f'}{f})^2\frac{f^{''}}{f}]
\end{eqnarray}
Where  we define in:
\begin{eqnarray}\nonumber
f=e^{-2\sqrt{\frac{\Lambda}{3}}r}(\xi^2e^{2\sqrt{3\Lambda}r}+1)^{2/3}
\end{eqnarray}
The poles of Kretschmann in real $r$ plane is determined by roots of
equation $f=0$. So we can conclude the following results\\
 1-If $\xi^2>0$ i.e. the scalar field's parameter is a real constant ,
 Then the Kretschmann scalar has no naked singularity at
 $r=r_{0}$.\\
 2- If we leave the above condition and allow  $\xi$ to be a
 complex value then the exact solution (14)  has  naked singularity at
 $r=r_{0}=-\frac{1}{3}a\ln(|\xi|)$ where $ |\xi|=Modulus(\xi)$. Only
 for $0<|\xi|<1$  this naked singularity is physically plausible.
If $|\xi|>1$  the exact solution  has no naked
 singularity.
Thus for our next  cosmological discussions that it is limited to
$\xi^2>0$ and real scalar fields , we can tell that our solution has
no singularity.
\section{Recovering previous solutions}
 In this section we show that our solution gives   $LC\Lambda$   and Buchdahl
 solutions.
 \subsection*{Case (1) :  limit of  the solution with $\Lambda\neq 0$,$\phi=constant$ ($LC\Lambda$ family)}
 We note that if $\xi\rightarrow 0$ the scalar field disappears and the Ricci
scalar (13) becomes $4\Lambda$ which is trivial in vaccum limit of
relation (2)\footnote{remind that this case trivially $\Box\phi=0$
satisfied and both signs $\pm $ in metric is admittable } . The line
element (1) in this case is read as:
\begin{eqnarray}
ds^{2}=dr^2+e^{\pm2\sqrt{\frac{\Lambda}{3}}r}(-dt^2+d\varphi^2+dz^2)
\end{eqnarray}
This space time is the other solution of General $LC\Lambda$ family
which has studied in \cite{5}.Because $r = 0 $ represents no special
hypersurface, we can also extend the range of r to
$(-\infty,\infty)$ and write the metric with the $+$ sign
only.Spacetime with metric (15)constitutes part of  de- Sitter
spacetime with metric written in the horospherical coordinates.
Solution (14) with the $+$ sign covers part of de -Sitter
hyperboloid with $x\in (0,\sqrt{\frac{3}{\Lambda}}]$
 while the $-$ sign solution is valid for $x\in
 [\sqrt{\frac{3}{\Lambda}},0)$ which
 $r=\pm\sqrt{\frac{3}{\Lambda}}\log(x\sqrt{\frac{\Lambda}{3}})$ in .
\subsection*{Case (2) :  limit of the  solution with $\Lambda=0$,$\phi\neq constant$(Buchdahl family)}
If $\Lambda\rightarrow 0$ we must at least recover one member of
Buchdahl cylindrical family. First we remind that from Buchdahl
theorem the general form of a massless minimally coupled scalar
field in axial symmetry has the following metric
form\footnote{Conventions on metric follow Buchdahl\cite{1}}:
\begin{eqnarray}
ds^2=e^{2(\gamma-\beta\psi)}(d\rho^2+dz^2)+\rho^2e^{-2\beta\psi}d\varphi^2-e^{2\beta\psi}dt^2
\end{eqnarray}
In which the metric functions $\psi,\gamma$ satisfy the following
equations:
\begin{eqnarray}
\rho^{-1}(\rho\psi_{,\rho})_{,\rho}+\psi_{,zz}=0\\
\gamma(\rho,z)=\int[(\rho\psi_{,\rho}^2-\psi_{,z}^2)d\rho+2\rho\psi_{,\rho}\psi_{,z}dz]\\
\phi=2\lambda\psi\\\beta=\pm(1-2\lambda^2)^{1/2}
\end{eqnarray}
If $\beta=1$ or $\lambda=0$ we obtain the seed vacuum solution in
axial symmetry, and if we focused only on cylindrical one , we would
have Levi-Civita solution \cite{8}. We must show that the metric
arisen from function (11) in limit $\Lambda\rightarrow 0$  belongs
to this family.The limit $\Lambda\rightarrow 0$ of the field (12)
must coincide with Buchdahl S-field for some value of
$\lambda,\psi$. Now we re derive  the general form of Buchdahl
metric in cylindrical symmetry. If the metric functions (16) depend
only on radial coordinate $\rho$ ,from equations (17,18) we obtain:
\begin{eqnarray}\nonumber
\gamma(\rho)=c_{1}^2\log(\rho)+c_{3}
\\\nonumber
\psi(\rho)=c_{1}\log(\rho)+c_{2}
\end{eqnarray}
After substituting these functions in (16) and comparing with metric
mass parameter, $m$ ,of Levi-Civita solution, also rewriting with
the non radial coordinates
$(t,\varphi,z)\rightarrow(\tilde{t},\tilde{\varphi},\tilde{z})$ we
finally have:
\begin{eqnarray}
ds^2=\rho^{2m(m-\beta)}(d\rho^2+d\tilde{z}^2)+\rho^{2-2\beta
m}d\tilde{\varphi}^2-\rho^{2\beta  m}d\tilde{t}^2
\end{eqnarray}
 Since we work on the gauge (1) , it is suitable to convert Buchdahl
 metric (21) to our gauge. So we apply the radial coordinate
 transformation :
\begin{eqnarray}\nonumber
\rho=[(m^2-m\beta+1)r]^{\frac{1}{m^2-m\beta+1}}
\end{eqnarray}
In term of this coordinate ,metric (21) transforms to:
\begin{eqnarray}
ds^2=dr^2-g_{tt}dt^2+g_{\varphi\varphi}d\varphi^2+g_{zz}dz^2
\end{eqnarray}
Which we have:
\begin{eqnarray}\nonumber
g_{tt}=[(m^2-m\beta+1)r]^{\frac{2m \beta}{m^2-m\beta+1}}\\\nonumber
g_{\varphi\varphi}=[(m^2-m\beta+1)r]^{\frac{2-2m
\beta}{m^2-m\beta+1}}\\\nonumber
g_{zz}=[(m^2-m\beta+1)r]^{\frac{2m(m- \beta)}{m^2-m\beta+1}}
\end{eqnarray}
Now we write the line element for (1) in limit
$\Lambda\rightarrow0$.Using field equations (5,6) in this limit and
supposing $\phi(r)\neq c.t.e$ (which leads to flat space as it is
shown previously) we can write the line element:
\begin{eqnarray}
ds^2=dr^2+[\frac{3}{2}(c_{1}r+c_{2})]^{\frac{2}{3}}(-dt^2+d\varphi^2+dz^2)
\end{eqnarray}
Without loss of generality we can take $c_{2}=0$. Comparing metric
(22) with (23), term by term, we obtain the following simple
algebraic equations:
\begin{eqnarray}\nonumber
\frac{3}{2}c_{1}=m^2-m\beta+1\\\nonumber m\beta=1\\\nonumber
m^2-4m\beta+1=0
\end{eqnarray}
 The solutions of these equations are:
\begin{eqnarray}\nonumber m=\pm\sqrt{3}\\\nonumber
\beta=\pm\frac{\sqrt{3}}{3}\\\nonumber c_{1}=2
\end{eqnarray}
By substituting of $\beta$ in (20) we have
$\lambda=\pm\frac{\sqrt{3}}{3}$. Thus we  show that our exact
solution gives at least one Buchdahl solution with proper values of
$\beta,m$.We note here that only in vaccum case $m=\pm1$ it
corresponds to flat space\cite{7} and in non vaccum case (as
Buchdahl family which contains Scalar field) it is a non flat
spacetime. Also if one can solve (3,4) completely we  expect that
the solution recovers all Buchdahl family with desired values of
$\beta,m$.
\section{Cosmological solutions}
 As we know , in General relativity , It is conceivable that some of
 the  solutions are related to cosmological solutions through a time-dependent scalar field
 with the cosmological constant.For an example we can treat a massless
minimally coupled scalar field in spherically symmetric space time
as a perfect fluid with equation of state $p=-\rho$ where  $p$ is
the principal pressure (radial pressure in spherical symmetry ) and
$\rho$ the energy density of scalar field enclosed  spacetime
\cite{16}. As i mentioned in \cite{3} this is a triviality in
spherical symmetry.There is no simple physical mechanism hidden
behind it. Many exact solutions have found describe perfect fluids
with a general equation of state $p=f(\rho)$. Taub and Tabensky show
that in plane symmetry, a perfect fluid solution can admit a scalar
field stress-energies \cite{12}. Our  solutions which is derived in
(8) can be converted to time-dependent metrics via the complex
transformation $ r\rightarrow i \tau$  and $ t\rightarrow i u$, as
 Vaidya and Som discussed in\cite{13}. Since the scale factor of the
resulting metric should be a real one valued function of time like
coordinate $w$ , the  physical , well defined and meaningful
solution is obtained by choosing the first linear solution of (5).As
we showed in case (1) this solution is a member of $LC\Lambda$
family.Another Cosmological solution can be obtained by choosing a
complex  transformation and having $\xi^2=1$ . This cosmology
describe as Case (2).\\
\subsection*{Case (1) : Cosmological model $\Lambda<0, \xi^2=0$}
Now we prove that under complex transformation this metric describe
static anti de Sitter space universe with a constant Hubble
parameter $H=\sqrt{\frac{-\Lambda}{3}}$. This case is occurred for a
negative Cosmological constant.Albeit this solution is in
contradiction to our intuition about the recent epoch of universe
which persuaded physicists to accept a  positive cosmological
constant, theoretically it's importance in relation to AdS
correspondence and  String Theory is momentous. If we apply  $
r\rightarrow i \tau$  and  $ t\rightarrow i u$ in - sign solution of
(15) we obtain:
\begin{eqnarray}
ds^{2}=-d\tau^2+e^{2\sqrt{\frac{-\Lambda}{3}}\tau}(du^2+d\varphi^2+dz^2)
\end{eqnarray}
The line element (24) is correspond to a Friedmann Cosmology with
Hubble parameter $H=\sqrt{\frac{-\Lambda}{3}}$ in the coordinates
$(\tau,u,\varphi,z)$. We can interpreted this solution as an
inflationary model where  the inflaton, whose energy density is
roughly constant, dominates the cosmic evolution and determines the
(roughly constant) Hubble parameter while causing the scale factor
to grow exponentially.
\subsection*{Case (2) : Cosmological model $\Lambda>0, \xi^2=1$}
The metric function in (14) must be a real function even in analytic
continuation of $t,r$ (complexified $t,r$).We observe that if we
choose the minus sign solution and set $\xi^2=1$ in (14) the metric
can then be converted to Robertson–Walker form:
\begin{eqnarray}
ds^{2}=-d\tau^2+(2\cos(\frac{3\tau}{a}))^{2/3}(du^2+d\varphi^2+dz^2)
\end{eqnarray}
 Where we define a de-Sitter  radius as
 $a=\sqrt{\frac{3}{\Lambda}}$.
This form of solution suggests  time coordinate $\tau$ as a periodic
coordinate.In the case of Cylindrical topology of the spatial
sections, $\tau,z$ and $\varphi$ are interpreted as generalized
Euler angle coordinates. Cyclic cosmology are based on the
braneworld picture of the universe \cite{15}, in where spacetime is
effectively 5-dimensional,with one dimension  which is not extending
indefinitely, but being a line segment. The cyclic model is an
ambitious attempt to provide a complete history of the universe,
within the framework of a braneworld view of the universe, while
incorporating both the ekpyrotic mechanism and dark energy in an
essential way. This solution repeated every once by a period
$T=\frac{2\pi a}{3}$.The metric (25) as a Freidmann model has a
bounded scale factor , i.e. for all epoches of the universe it
satisfies the following inequality(in natural units),
\begin{eqnarray}\nonumber
a(\tau)\leq 1.2599
\end{eqnarray}

\section{Summary and conclusions}
Exact solution of Einstein theory is discussed in many different
Symmetries with a diversified class of stress-energies. For a scalar
field there is only a few exact solutions in spherical, cylindrical
or plane symmetries. In this manner the generating methods are so
important. one of the simple ones which is applicable for a massless
scalar field is Buchdahl Reciprocal method that if we applied it to
a static vaccum solution of Einstein equations in cylindrical
symmetry we would obtain a one parameter exact solution which
recovers seed metric (LC spacetime) in a special case. In this
article we generalized Buchdahl metric to that one with a
cosmological constant term and by considering more general
cylindrical metric in Weyl gauge , we found a new 2 parameter  exact
solution for a massless scalar field in the presence of the
cosmological constant term. Far from the importance of this solution
as a new exact solution , this solution can be interpreted as a
cosmological solution by complexities radial and time coordinates.
Two distinct classes of solutions to the problem of a
cylindrically-symmetric static field in general relativity with
cosmological constant have been found. The first one of these
solutions involves two parameters, cosmological constant term
$\Lambda$  and $\xi$  , which is related to a scalar field. Various
choices of $\xi$ demonstrated that the presented solution was a
generalization of the metrics previously derived by Buchdahl , Linet
\cite{10} and Tian \cite{11}. These solutions are static Cosmic
string spacetimes. Another solution  contains flat space . The limit
of solution in the absence of scalar field coincides with a number
of $LC\Lambda$ family which has many attractive properties . As an
example, Spacetime with
 this metric constitutes part of  de- Sitter spacetime with metric which is
written in the horospherical coordinates. We also present two
cosmological models by analytic continuation the time and radial
coordinates. First one corresponds to an inflationary model that
 the Hubble parameter remains constant in it and has no dynamics.
Another Cosmological solution is a cosmic time periodic model which
can be classified as a Cyclic model which has some important
applications in braneworld scenario.

\section{Acknowledgement}
D.Momeni thanks  Anzhong Wang for useful comments and valuable
suggestions.

\end{document}